\documentclass{tise_style}\usepackage[]{graphicx}\usepackage[]{color}
\makeatletter
\def\maxwidth{ %
  \ifdim\Gin@nat@width>\linewidth
    \linewidth
  \else
    \Gin@nat@width
  \fi
}
\makeatother

\definecolor{fgcolor}{rgb}{0.345, 0.345, 0.345}
\newcommand{\hlnum}[1]{\textcolor[rgb]{0.686,0.059,0.569}{#1}}%
\newcommand{\hlstr}[1]{\textcolor[rgb]{0.192,0.494,0.8}{#1}}%
\newcommand{\hlopt}[1]{\textcolor[rgb]{0,0,0}{#1}}%
\newcommand{\hlstd}[1]{\textcolor[rgb]{0.345,0.345,0.345}{#1}}%
\newcommand{\hlkwb}[1]{\textcolor[rgb]{0.69,0.353,0.396}{#1}}%
\newcommand{\hlkwc}[1]{\textcolor[rgb]{0.333,0.667,0.333}{#1}}%
\newcommand{\hlkwd}[1]{\textcolor[rgb]{0.737,0.353,0.396}{\textbf{#1}}}%

\usepackage{framed}
\makeatletter
\newenvironment{kframe}{%
 \def\at@end@of@kframe{}%
 \ifinner\ifhmode%
  \def\at@end@of@kframe{\end{minipage}}%
  \begin{minipage}{\columnwidth}%
 \fi\fi%
 \def\FrameCommand##1{\hskip\@totalleftmargin \hskip-\fboxsep
 \colorbox{shadecolor}{##1}\hskip-\fboxsep
     \hskip-\linewidth \hskip-\@totalleftmargin \hskip\columnwidth}%
 \MakeFramed {\advance\hsize-\width
   \@totalleftmargin\z@ \linewidth\hsize
   \@setminipage}}%
 {\par\unskip\endMakeFramed%
 \at@end@of@kframe}
\makeatother

\definecolor{shadecolor}{rgb}{.97, .97, .97}
\definecolor{messagecolor}{rgb}{0, 0, 0}
\definecolor{warningcolor}{rgb}{1, 0, 1}
\definecolor{errorcolor}{rgb}{1, 0, 0}
\newenvironment{knitrout}{}{} 

\usepackage{alltt}
\usepackage{url}
\PassOptionsToPackage{hyphens}{url} 
\usepackage{graphicx}
\usepackage{float}
\usepackage{verbatim}
\usepackage{fullpage}
\usepackage{color}
\usepackage{natbib}
\usepackage{caption}
    \title{ Dynamic Data in the Statistics Classroom}

    \author{Johanna Hardin}
\IfFileExists{upquote.sty}{\usepackage{upquote}}{}
\begin{document}
\maketitle



\section{{ Introduction}}

There has been a recent push to change the way we - as statisticians - engage pedagogically with complex real-world data analysis problems.   Two parallel forces have directed us toward embracing more complex-data real-world problems in the classroom.

The first driver is a call from educators to make statistics more relevant to students' experiences.  In order to address the perspective of students who previously have been reported to ``exhibit remarkably little curiosity about the material they are analyzing", \citet{brow:kass:2009} suggest the importance of ``real-world problem solving" to get students engaged in their analyses.  \citet{goul:2010} wants introductory statistics students to leave the course with ``a set of ... attitudes about data that are immediately applicable to their lives."  \citet{goul:cent:2013} suggest putting ``data at the center of the curriculum.''  \citet{hort:baum:2015} assert that ``statistics students need to develop the capacity to make sense of the staggering amount of information collected in our increasingly data-centered world.''  And \citet{zhu:2013} remind us that ``data pre-processing bridges the gap from data acquisition to statistical analysis but has not been championed as a relevant component in statistics curricula.''

The second driver arises from students and other stakeholders and is harder to document with references.   Certainly my own experience (and that of my colleagues) is that students engage at the deepest level when they (a) care about the problem, and (b) understand the data collection, study design, or motivation of the problem.  \citet{stur:kuip:2015} report,  ``Our student evaluations of these materials support Gould's \citeyearpar{goul:2010} comments suggesting that previously collected and cleaned data were considered abstract to the student ... We have found that unless students 1) have collected the data themselves or 2) clearly see where and how the data were collected, they often fail to appreciate it."

Indeed, even if the work is done for them, as \citet{grim:2015} argues, there is much benefit in the students seeing how the data were procured:
\begin{quote}
Using the vocabulary of \citet{wick:2014}, teachers hide the `messy data' aspects and provide `tidy data' - even when students possess the data skills required to work with the messy data. It is valuable for students to not only have many authentic data experiences but also to have the professor model the correct application of statistics by showing work with messy data in lectures...What is good for statistics majors can also be applied to introductory courses. There may be no data skills on the learning outcomes for these courses, but some examples and homework in an introductory course may be modified and/or updated to use the original source data instead of a curated dataset. The data skills required would certainly be modest and need to fit student backgrounds. The objective would be for students to see that data skills are required in an analysis. Students may rely on code provided to them that results in their own copy of the dataset.
\end{quote}

There have been repeated calls to use {\em real data} in the classroom \citep{cobb:1991,cobb:1992,cobb:2007,cobb:2011,asa-undergrad,asa:2014,gaise,gaise2015}.  There are benefits to using real data sets  (as opposed to made up numbers) in textbooks and in the classroom.  Indeed, there are virtually no textbooks of any kind (AP Statistics, Introductory Statistics, second course in statistics) being written today without the vast majority of examples taken from actual studies or databases.  Additionally, there has been a push to infuse R with datasets that are relevant, recent, and sophisticated (e.g., nycflights13 and  other mosaic datasets, \citep{mosaic}).

Unfortunately, by nature, however, all of the data given in a textbook or an R package is static.  Currently, there do not exist mechanisms to continually update any dataset provided by the course materials.  The most comprehensive baseball dataset compiled and provided to the students will be out of date (and less interesting) by next fall.  The good news, however, is that, outside of the statistics classroom, data of all kinds are being updated in real time publicly and accessibly.  And even better news is that R developers are continually improving interfaces to the vast amounts of public data.  For example, the \verb;tidyverse; package imports a handful of packages that make downloading data straightforward (e.g., see \citet{wick:2014} and more recently the packages: \verb;rvest;, \verb;readr;, \verb;readxl;, \verb;haven;, \verb;httr;, and \verb;xml2;; expect more to come).

\citet{grim:2015} delineates data along two axes:  the first axis describes the source and format for the data; the second axis describes the amount of wrangling required on the data.  Both axes are scored as good/better/best.    Indeed, Grimshaw gives examples, sources, and R code for several examples.  However, he included only one dataset from the ``best/best" category that worked well in his classroom.  Out of the three ``best/best" datasets he used, two were not well received because of excessive time, computing expertise, or domain knowledge needed to wrangle the data into usable format and extract meaning from the dataset.

In this manuscript, we seek to address the difficulty of using dynamic data in the classroom by curating additional best/best datasets (including full R Markdown files to facilitate reproducible analysis).  The available resources aim to open the world of dynamic data to those who have not previously worked directly with downloaded data.  We hope our work can act as a starting point for those interested in building their data scraping skills.  Additionally, the R Markdown files can be used as scaffolding for assignments designed to help students engage with online resources.

In the next section we work through an entire example including showing the steps and R code for downloading the data, providing example code for using it in a typical introductory statistics classroom, and suggesting ideas for expanded use beyond introductory statistics.   Section 3 gives summaries of additional dynamic datasets that are fully curated as R Markdown files in the supplementary resources on Github (\url{https://github.com/hardin47/DynamicData}) as well as a list of other places to look for dynamic data, and we provide closing thoughts and ideas for future work in the Conclusion.  The Appendix describes a few technical details that are helpful for making dynamic data accessible to students at all levels.

\section{Complete Example: College Scorecard Data}

Supplementary materials provide full R Markdown files for complete analysis on five different dynamic datasets.  We start with a complete description of the College Scorecard materials as a way of illustrating the available resources.

Data on characteristics of US institutions of higher education were collected in an effort to make more transparent issues of cost, debt, completion rates, and post-graduation earning potential.  An undertaking of the U.S. Department of Education, the College Scorecard data represent a compilation of institutional reporting, federal financial aid reports, and tax information.  The process of gathering and compiling the data is well documented on the College Scorecard website \url{https://collegescorecard.ed.gov/data/documentation/}.  One caveat is that some of the variables have only been collected on students receiving federal financial aid.  Biases inherent to analyses done on data collected from a subgroup should be considered.

The College Scorecard dataset is incredibly rich.  The individual institutions are broken down by total share of enrollment of various races, family income levels, first generation status, age of students, etc.  Additionally, the data give matriculation information like SAT scores as well as graduation information like completion rate and income level.  The dataset allows for a student to investigate political or personal hypotheses about college education and the costs and benefits within.  The variables are described in a data dictionary given at \url{https://collegescorecard.ed.gov/assets/CollegeScorecardDataDictionary.xlsx}.

\subsection{Downloading the Data}

For each of the five fully curated dynamic datasets, there is an R Markdown file (available at \url{https://github.com/hardin47/DynamicData}) which scrapes the data from an outside web source (presumably kept current and public by some other organization).   The downloading step is shown here in the manuscript only for the College Scorecard data.  Note, as discussed above, the data are downloaded directly from the website managed by the US Department of Education (and not stored locally as a csv file).  It is worth pointing out that the code given here is more complicated than what is standard in an introductory statistics course (especially at the beginning of the semester!).  However, both the GapMinder and Wikipedia examples can be adjusted in very simple ways, allowing each student to work with their own dataset.  That is, you can scaffold an assignment by providing the majority of the downloading code and having the student fill in the URL (Wikipedia example) or variable names (GapMinder example).  The R code for downloading the data for all of the examples are given in the supplementary R Markdown files, and each uses slightly different functions and syntax.

First, load the data into R.
\begin{knitrout}\footnotesize
\definecolor{shadecolor}{rgb}{0.969, 0.969, 0.969}\color{fgcolor}\begin{kframe}
\begin{alltt}
\hlstd{college_url} \hlkwb{<-}
  \hlstr{"https://s3.amazonaws.com/ed-college-choice-public/Most+Recent+Cohorts+(All+Data+Elements).csv"}
\end{alltt}
\end{kframe}
\end{knitrout}

\begin{knitrout}
\definecolor{shadecolor}{rgb}{0.969, 0.969, 0.969}\color{fgcolor}\begin{kframe}
\begin{alltt}
\hlstd{college_data} \hlkwb{<-} \hlstd{readr}\hlopt{::}\hlkwd{read_csv}\hlstd{(college_url)}
\hlkwd{dim}\hlstd{(college_data)}
\end{alltt}
\begin{verbatim}
## [1] 7804 1728
\end{verbatim}
\end{kframe}
\end{knitrout}

Next, use data wrangling methods to clean and organize the variables.
\begin{knitrout}
\definecolor{shadecolor}{rgb}{0.969, 0.969, 0.969}\color{fgcolor}\begin{kframe}
\begin{alltt}
\hlstd{college_debt} \hlkwb{=} \hlstd{college_data} \hlopt{%>%}
  \hlstd{dplyr}\hlopt{::}\hlkwd{select}\hlstd{(region, HBCU, DEBT_MDN, md_earn_wne_p10)} \hlopt{%>%}
  \hlkwd{mutate}\hlstd{(}\hlkwc{DEBT_MDN} \hlstd{= readr}\hlopt{::}\hlkwd{parse_number}\hlstd{(DEBT_MDN),}
         \hlkwc{md_earn_wne_p10} \hlstd{= readr}\hlopt{::}\hlkwd{parse_number}\hlstd{(md_earn_wne_p10))} \hlopt{%>%}
  \hlkwd{mutate}\hlstd{(}\hlkwc{HBCU} \hlstd{=} \hlkwd{ifelse}\hlstd{(HBCU}\hlopt{==}\hlstr{"NULL"}\hlstd{,} \hlnum{NA}\hlstd{, HBCU))} \hlopt{%>%}
  \hlkwd{mutate}\hlstd{(}\hlkwc{region2} \hlstd{=} \hlkwd{ifelse}\hlstd{(region}\hlopt{==}\hlstr{"0"}\hlstd{,} \hlstr{"Military"}\hlstd{,}
                  \hlkwd{ifelse}\hlstd{(region}\hlopt{==}\hlstr{"1"}\hlstd{,} \hlstr{"New England"}\hlstd{,}
                  \hlkwd{ifelse}\hlstd{(region}\hlopt{==}\hlstr{"2"}\hlstd{,} \hlstr{"Mid East"}\hlstd{,}
                  \hlkwd{ifelse}\hlstd{(region}\hlopt{==}\hlstr{"3"}\hlstd{,} \hlstr{"Great Lakes"}\hlstd{,}
                  \hlkwd{ifelse}\hlstd{(region}\hlopt{==}\hlstr{"4"}\hlstd{,} \hlstr{"Plains"}\hlstd{,}
                  \hlkwd{ifelse}\hlstd{(region}\hlopt{==}\hlstr{"5"}\hlstd{,} \hlstr{"Southeast"}\hlstd{,}
                  \hlkwd{ifelse}\hlstd{(region}\hlopt{==}\hlstr{"6"}\hlstd{,} \hlstr{"Southwest"}\hlstd{,}
                  \hlkwd{ifelse}\hlstd{(region}\hlopt{==}\hlstr{"7"}\hlstd{,} \hlstr{"Rocky Mnts"}\hlstd{,}
                  \hlkwd{ifelse}\hlstd{(region}\hlopt{==}\hlstr{"8"}\hlstd{,} \hlstr{"Far West"}\hlstd{,} \hlstr{"Outlying"}\hlstd{))))))))))}

\hlkwd{summary}\hlstd{(college_debt)}
\end{alltt}
\begin{verbatim}
##      region          HBCU              DEBT_MDN      md_earn_wne_p10 
##  Min.   :0.000   Length:7804        Min.   :   333   Min.   :  8400  
##  1st Qu.:3.000   Class :character   1st Qu.:  7710   1st Qu.: 24200  
##  Median :5.000   Mode  :character   Median :  9833   Median : 31200  
##  Mean   :4.621                      Mean   : 11830   Mean   : 33233  
##  3rd Qu.:6.000                      3rd Qu.: 15462   3rd Qu.: 39200  
##  Max.   :9.000                      Max.   :131335   Max.   :250000  
##                                     NA's   :1163     NA's   :2168    
##    region2         
##  Length:7804       
##  Class :character  
##  Mode  :character  
##                    
##                    
##                    
## 
\end{verbatim}
\end{kframe}
\end{knitrout}

\subsection{Using dynamic data within a typical classroom}
Using the downloaded data, we start by applying a technique from the introductory curriculum to a research question of interest based on the College Scorecard data.  College debt is of particular interest to many college students, but debt can be mediated by post-graduation income.  To fully investigate the relationship between the variables, we provide both confidence and prediction intervals for both variables.  

After calculating a few individual intervals, we show all intervals represented graphically and broken down by geographic region. Note that the visual representations do not represent a simple summary plot of the data, and we leave it open to the instructor to have the students engage more deeply with the many available variables.

Using the two variables measuring amount of debt of a typical (i.e., median) college graduate and median earning 10 years after matriculation, we create both confidence intervals and prediction intervals -- keeping in mind that the observational unit is an academic institution.  Note that the calculations below are for both confidence and prediction intervals.  The confidence interval agglomerates institutions over the entire dataset; however, the prediction value is for a single {\em institution}  (which is the observational unit).  The analysis lends itself nicely to a conversation about confidence vs. prediction intervals as well as observational units as institution vs. as individual student.  It is worth pointing out to the students that the prediction intervals likely hold more information related to their individual experiences than the confidence intervals.  However, the unit of prediction is for an {\em institution}, and so the individual student debt and income is likely even more variable than shown here. Additionally, Figure \ref{worth} demonstrates the effect of samples size:  consider the comparison of the Military intervals (one school) to the intervals for all of the US institutions (about 6000 schools).  (Note: the intervals given in Figure \ref{worth} were created using an ANOVA model where the within variance is calculated across all regions, which is how the interval for military schools can be calculated.  You may or may not want to bring that up with your students.)

The following R code uses the \verb;mosaic; package to directly calculate both prediction and confidence intervals.  Note the formula interface given by the tilde is described in detail here: \url{http://rpruim.github.io/eCOTS2014/Workshop/Modeling.html}.

\begin{knitrout}
\definecolor{shadecolor}{rgb}{0.969, 0.969, 0.969}\color{fgcolor}\begin{kframe}
\begin{alltt}
\hlstd{debt_mod} \hlkwb{<-} \hlkwd{lm}\hlstd{(DEBT_MDN}\hlopt{~}\hlnum{1}\hlstd{,} \hlkwc{data} \hlstd{= college_debt)}
\hlstd{debt_fun} \hlkwb{<-} \hlstd{mosaic}\hlopt{::}\hlkwd{makeFun}\hlstd{(debt_mod)}
\hlkwd{debt_fun}\hlstd{()}
\end{alltt}
\begin{verbatim}
##        1 
## 11829.78
\end{verbatim}
\begin{alltt}
\hlkwd{debt_fun}\hlstd{(}\hlkwc{interval}\hlstd{=}\hlstr{"confidence"}\hlstd{)}
\end{alltt}
\begin{verbatim}
##        fit      lwr      upr
## 1 11829.78 11692.44 11967.13
\end{verbatim}
\begin{alltt}
\hlkwd{debt_fun}\hlstd{(}\hlkwc{interval}\hlstd{=}\hlstr{"prediction"}\hlstd{)}
\end{alltt}
\begin{verbatim}
##        fit      lwr      upr
## 1 11829.78 636.2383 23023.33
\end{verbatim}
\begin{alltt}
\hlstd{earn_mod} \hlkwb{<-} \hlkwd{lm}\hlstd{(md_earn_wne_p10}\hlopt{~}\hlnum{1}\hlstd{,} \hlkwc{data} \hlstd{= college_debt)}
\hlstd{earn_fun} \hlkwb{<-} \hlstd{mosaic}\hlopt{::}\hlkwd{makeFun}\hlstd{(earn_mod)}
\hlkwd{earn_fun}\hlstd{()}
\end{alltt}
\begin{verbatim}
##        1 
## 33232.59
\end{verbatim}
\begin{alltt}
\hlkwd{earn_fun}\hlstd{(}\hlkwc{interval}\hlstd{=}\hlstr{"confidence"}\hlstd{)}
\end{alltt}
\begin{verbatim}
##        fit      lwr      upr
## 1 33232.59 32864.78 33600.41
\end{verbatim}
\begin{alltt}
\hlkwd{earn_fun}\hlstd{(}\hlkwc{interval}\hlstd{=}\hlstr{"prediction"}\hlstd{)}
\end{alltt}
\begin{verbatim}
##        fit      lwr      upr
## 1 33232.59 5616.893 60848.29
\end{verbatim}
\end{kframe}
\end{knitrout}

The intervals are interesting, but they might be even more interesting if broken down by region and shown visually.  Note how much smaller the confidence intervals are from the prediction intervals!  The difference indicates lots of variability across institutions and large sample sizes.

\begin{figure}[H]
\begin{center}
\begin{knitrout}
\definecolor{shadecolor}{rgb}{0.969, 0.969, 0.969}\color{fgcolor}
\includegraphics[width=\maxwidth]{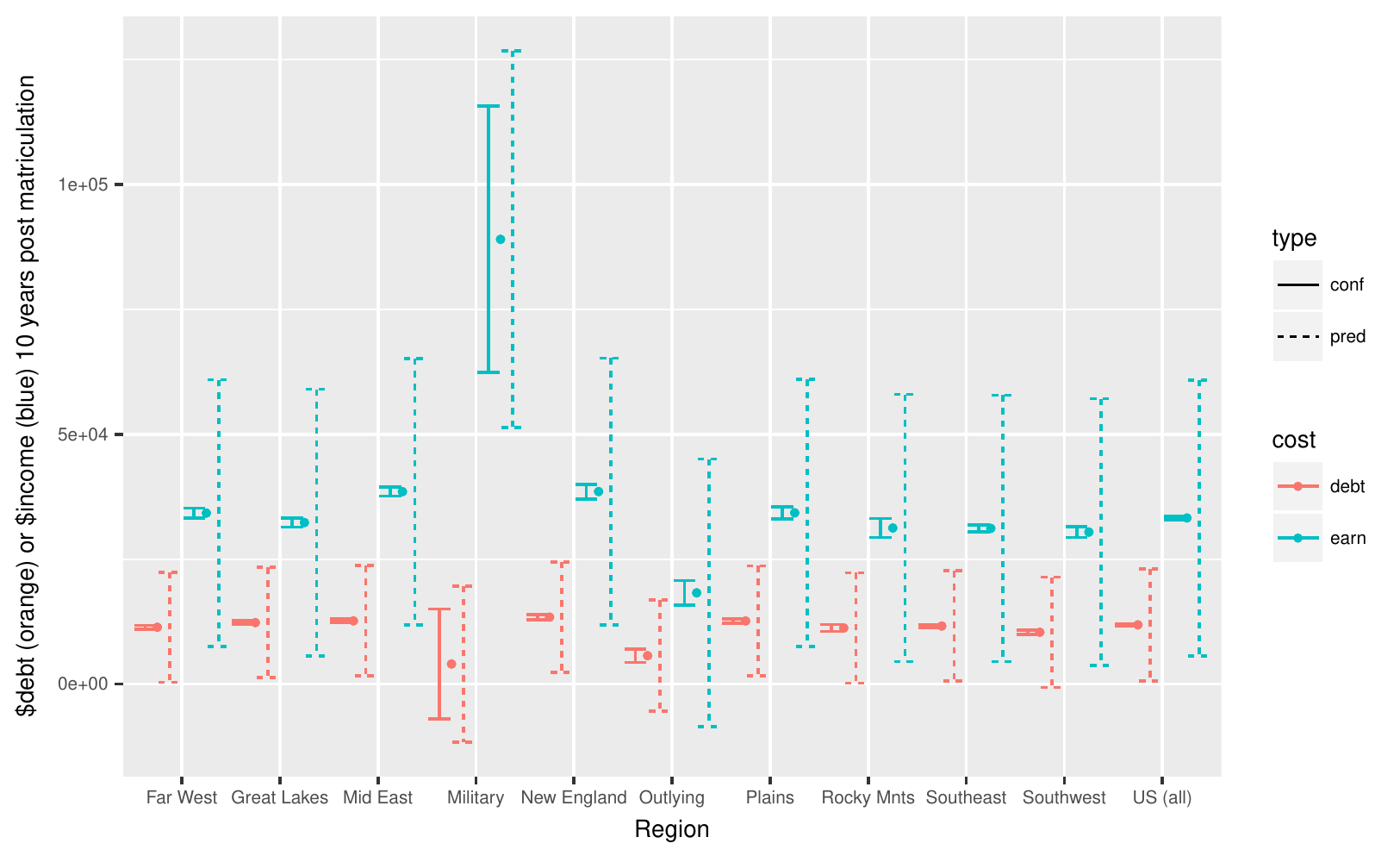} 

\end{knitrout}
\caption{\label{worth}  \footnotesize The x-axis represents the region of the institution.  The y-axis represents either the amount of debt 10 years after matriculation (orange) or the amount of income 10 years after matriculation (blue).  Confidence intervals for the average values (within region) are given by the solid lines.  Prediction intervals for individual institutions are given by the dashed lines.  The solid dot represents the center of both types of intervals (broken down by debt and income).}
\end{center}
\end{figure}

\subsection{Thinking outside the box}
Additionally, for each of the supplementary dynamic data R Markdown files, we add a section for each analysis which is based on topics that are not traditionally taught in introductory statistics classes.  The additional analysis is done not only to expand the tool box of the students but also to teach the students that they can often think about the problem in sophisticated ways even if all their tools come only from the introductory course.

\begin{figure}[H]
\begin{center}
\begin{knitrout}
\definecolor{shadecolor}{rgb}{0.969, 0.969, 0.969}\color{fgcolor}
\includegraphics[width=\maxwidth]{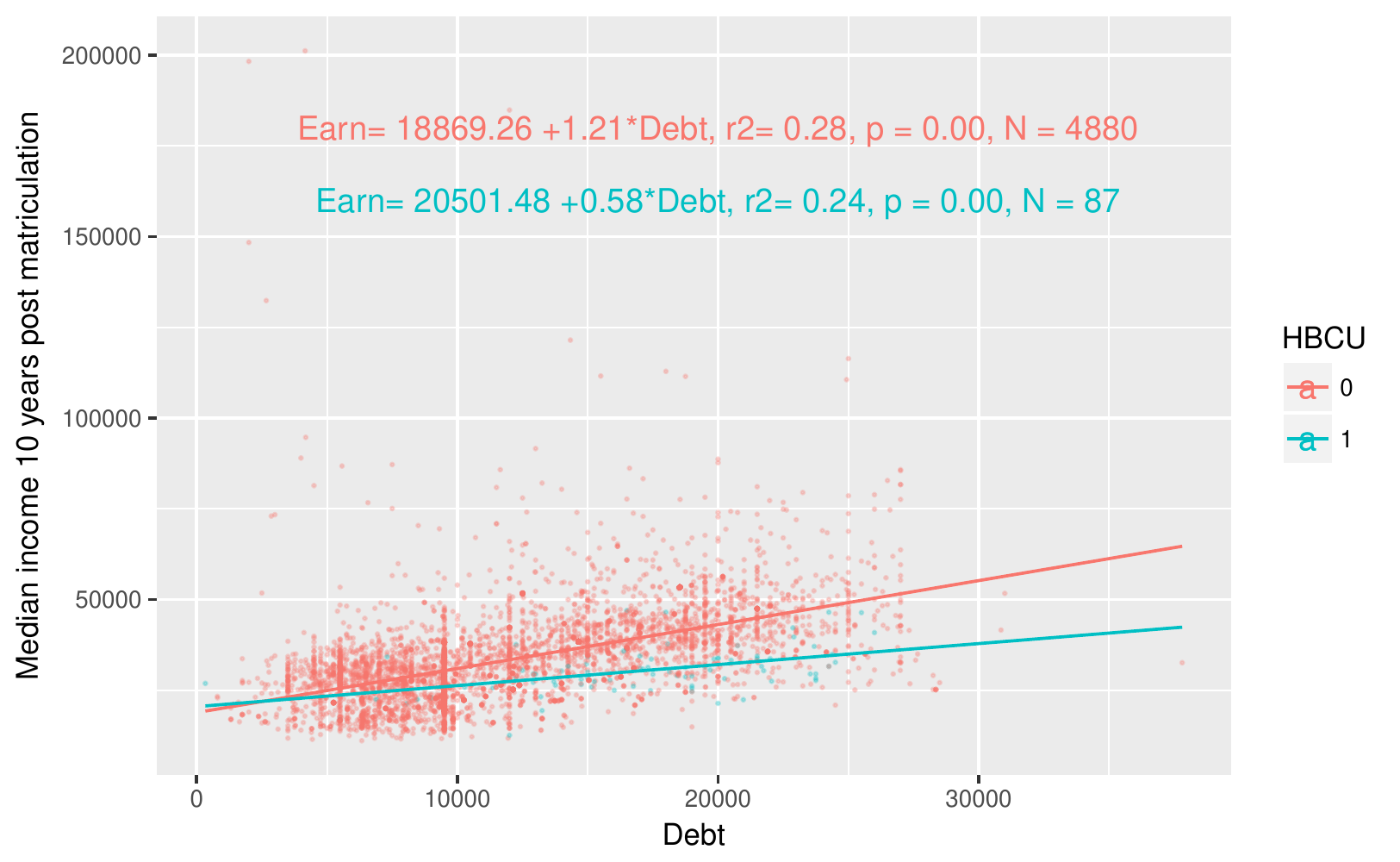} 

\end{knitrout}
\caption{\label{hbcu} \footnotesize Median income regressed on debt.  For the analysis, HBCU is interacted with debt to provide two distinct (and not parallel) regression lines.  HBCU institutions are given in blue, and non-HBCU institutions are given in orange.}
\end{center}
\end{figure}

The College Scorecard dataset is incredibly rich and can be used for many different types of model building: linear, logistic, machine learning.  Indeed, thinking about interaction terms could be particularly insightful. Here, we give an example of regressing earnings on debt with the interaction term as whether or not the institution is one of the Historically Black Colleges and Universities (HBCU). Figure \ref{hbcu} displays the separate regression lines for the two distinct types of institutions.

Many interesting conversations can ensue based on the regression of income on debt.  Reminding the students that each observation is an institution is an important starting point.  Additionally, students should be able to volunteer the dangers of using a model like this to suggest causality.  Last, there might be room to discuss an inferential analysis of whether HBCUs are statistically different from non-HBCUs (noting the substantial differences in sample sizes).

It is not hard to come up with additional questions to investigate with the College Scorecard data.  Indeed, because the data relate directly to college students, they should be able to find ways to engage with the data.  We recommend continued conversations about how the data are valuable to the larger community, but that the information is not always complete (e.g., many variables are collected only on students who fill out financial aid forms) and not causative.

\section{Dynamic Data Projects}

For each of four additional dynamic datasets, we describe the source of the data and relevant variables \& research questions, some standard and graphical techniques (``using dynamic data within a typical introductory statistics classroom"), and a statistical analysis appropriate for a course after introductory statistics (``thinking outside the box").  We also provide source information for an additional nine dynamic datasets.

\subsection{Wikipedia Data}

Wikipedia stores most of its tabular data in HTML tables.  To scrape HTML tables from any website (or HTML file), use the R function {\em XML::readHTMLTable}.  As an initial foray into downloading data directly from the internet into R, Wikipedia tables provide a nice introduction.  In the supplementary R Markdown file associated with the Wikipedia data analysis, we also walk through some of the useful aspects of using \verb;dplyr; to wrangle the data\footnote{Original idea for this example provided by Nick Horton, Amherst College.}.

\begin{figure}[H]
\begin{center}
\begin{knitrout}
\definecolor{shadecolor}{rgb}{0.969, 0.969, 0.969}\color{fgcolor}
\includegraphics[width=\maxwidth]{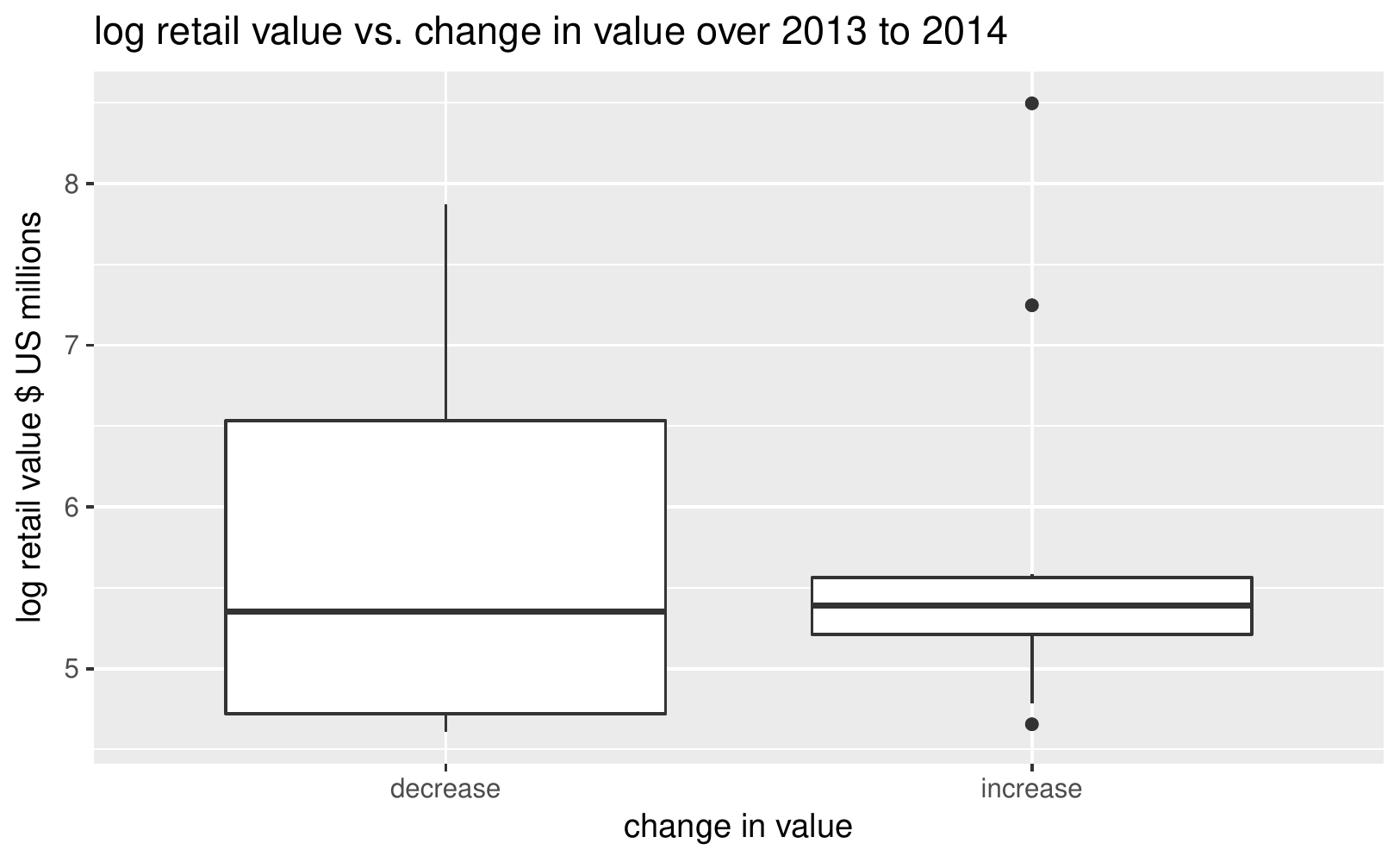} 

\end{knitrout}
\caption{\label{musicbox}  \footnotesize  The log of the retail value of music sales broken down by whether or not the sales have increased or decreased (presumably over the previous year, although the Wikipedia documentation does not specify the time period over which the change is measured).}
\end{center}
\end{figure}

\vspace*{-.5cm}
\subsubsection*{Using dynamic data within a typical classroom}
The Wikipedia analysis given in the fully curated files explores an HTML table on sales of music (physical and digital) in 2014, \url{https://en.wikipedia.org/wiki/Music_industry}.  One variable gives an indication of whether the retail value of the music sales has increased or decreased.  Using the country-level music data, we perform a t-test, a Wilcoxon rank sum test, data transformations, and boxplots to investigate music retail sales (analysis given in supplementary materials, not shown here).  By grouping the data into two categories we can investigate whether there is any statistical difference between the total average retail sales (in US\$) between those countries for whom retail sales increased versus those that decreased.  The p-value for the initial t-test is reasonably large, but the boxplot shows that the difference in variability across the two groups is also large with a sample that either has large outliers or a long skewed right tail.  Because the technical assumptions do not appear to be met, a log transformation of the data or a non-parametric test might be better assessments of the data (see Figure \ref{musicbox}).   The analysis leads to conversations about the source of the data and the reasons why p-values are non-significant.  The example extends easily to each student choosing their own Wikipedia page and data table, graphical representations, and statistical analyses.

\subsubsection*{Thinking outside the box}

Among the variables in the Wikipedia music dataset are the breakdown (percentages) of how the retail sales are distributed across physical, digital, performance rights, and synchronization.  We might want to see whether there is a dependency of total retail sales on the breakdown of types of products.  The problem is not well suited to introductory statistics as there is not an obvious statistic we can use within a sampling distribution (to create a p-value, etc.).  Because there does not seem to be an obvious mechanism for evaluating the breakdown of products (and how ``different" they are), we consider an ad-hoc measure and perform a permutation test to assess significance.  The average breakdown of retail sales is given in the R Markdown file in the supplementary materials.  One way to measure a discrepancy between the retail sales and the consistency of product breakdown is to correlate the retail sales with the sum of squared distances from the average breakdown of product types.  We see that the metric we created to find a relationship between retail sales and breakdown of types of product does not show significance.  A student project could be to think about different ways to measure how the breakdowns can be considered to be different.

\subsection{NHANES Data}

The NHANES data come from the National Health and Nutrition Examination Survey, surveys given nationwide by the Center for Disease Controls (CDC).  The CDC adopted the following sampling procedure:

\begin{enumerate}
\item Selection of primary sampling units (PSUs), which are counties or small groups of contiguous counties.
\item Selection of segments within PSUs that constitute a block or group of blocks containing a cluster of households.
\item Selection of specific households within segments.
\item Selection of individuals within a household
\end{enumerate}

About 12,000 persons per 2-year cycle were asked to participate in NHANES. Response rates varied by year, but an average of 10,500 persons out of the initial 12,000 agreed to complete a household interview. Of these, about 10,000 then participated in data collection at the mobile exam center. The persons (observational units) are located in counties across the country. About 30 selected counties were visited during a 2-year survey cycle out of approximately 3,000 counties in the United States. Each of the four regions of the United States and metropolitan and non-metropolitan areas are represented each year.  As such, the data collection is ongoing, and the data are updated on the NHANES website periodically, \url{http://www.cdc.gov/nchs/nhanes.htm}.  This manuscript uses the 2011-2012 NHANES data, but we expect the data to be updated regularly, and the URL should simply change to 2013-2014 when it becomes available.

Note that the NHANES data are available in the \verb;mosaic; package \citep{mosaic}, but the \verb;mosaic; version of the NHANES data is static (from 2011-2012), and the data has been cleaned with pre-selected variables.  Additionally, the variables can be downloaded directly using the \verb;nhanesA; package in R.  \url{https://cran.r-project.org/web/packages/nhanesA/vignettes/Introducing_nhanesA.html}.  By accessing the data directly from the CDC's website, students become more involved in the data analysis process, understanding what they can and cannot get from the data.  

The variables in the CDC's online NHANES dataset are virtually limitless.  We use a few different datasets, merging them based on an individual identifier in the dataset.  The variable information is all given online, but each at a different webpage.  For example, the demographic data is at \url{http://wwwn.cdc.gov/nchs/nhanes/search/variablelist.aspx?Component=Demographics&CycleBeginYear=2011}.

A further important aspect to the example is that (as described above) the data do not constitute a simple random sample (it is a weighted sample) from a population.  The sampling scheme can be part of the statistical inquiry into the data analysis, or it can be set by the instructor in the template used to download the data.  The data which is directly downloaded from the CDC website includes variables on the weighting scheme.  In the R Markdown file provided with this manuscript, we demonstrate how to create a dataset which {\em can} act as a simple random sample from the population.\footnote{The weighting analysis was motivated by work done by Shonda Kuiper (\url{http://web.grinnell.edu/individuals/kuipers/stat2labs/weights.html}) as well as the Project Mosaic Team (\url{https://cran.r-project.org/web/packages/NHANES/NHANES.pdf}).}  The figure and analysis below are done with a proxy simple random sample.

\subsubsection*{Using dynamic data within a typical classroom}

In the supplementary materials (not shown here), we start with a comparison of body mass index (BMI) for those in committed relationships and those not in committed relationships.  The graphs of the two BMI distributions look quite similar, and the t-test shows a non-significant difference in means.  The results prompt a discussion about averages versus individual results, causation, and sample size.

\subsubsection*{Thinking outside the box}

\begin{figure}[H]
\begin{center}
\begin{knitrout}
\definecolor{shadecolor}{rgb}{0.969, 0.969, 0.969}\color{fgcolor}
\includegraphics[width=\maxwidth]{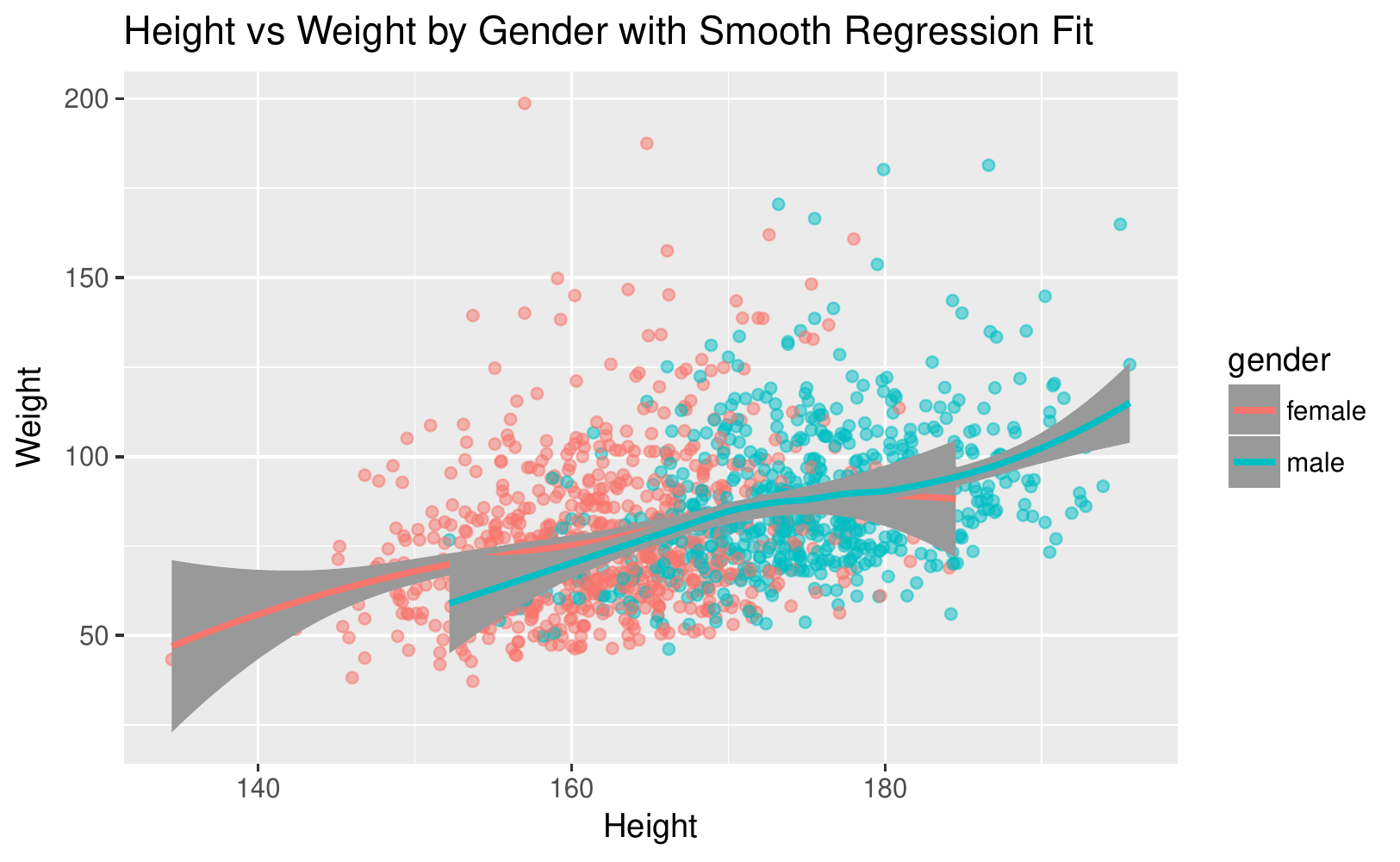} 

\end{knitrout}
\caption{\label{nhanesSmooth} \footnotesize A scatterplot of weight and height broken down by gender.  A smooth regression is fit to the points, and the confidence interval for the smooth curve as well as the increase in variability for small and large values of height are part of an important discussion in a second level regression course.}
\end{center}
\end{figure}

Because of the large sample size and the ability to determine the functional form of a non-linear relationship, smoothing techniques can be used to model quantitative variables.  Adding a smooth curve to a standard scatterplot leads (see Figure \ref{nhanesSmooth}) to discussions about how smooth curves are estimated, SE of the smooth curve, extra variability and instability due to extremes and fewer data points on the ends.  However, extrapolation (note that the two curves have different ranges?) and the slopes of the two curves not seeming to be different (no interaction?) might warrant further study.

\subsection{Global Literacy Rate Data}
The next dynamic dataset comes from GapMinder \url{http://www.gapminder.org/}.  GapMinder also has a plethora of variables from which students can choose (according to their own interests), but here we work with literacy rates measured at the country level. The analysis (and more importantly, the data scraping from GapMinder) could easily be extended for students interested in all sorts of political, social, environmental, or demographic data available.  Understanding political and demographic trends across both time and location can provide very interesting insight into economic or political science questions.  Alternatively the GapMinder data can be perused in a descriptive or graphical manner.

\subsubsection*{Using dynamic data within a typical classroom}

The introductory analysis considers gender differences in literacy rates and uses a linear model on the difference between female and male literacy rates across time (the analysis is available in the supplementary materials and not shown here).  We show a graphical representation and discuss model assumptions including sampling and independence of residuals. The model indicates that the difference between male and female literacy rates is shrinking over time.  However, we worry about the effects of other variables and encourage a more complete analysis.  Indeed, there may be large biases in our model if important explanatory variables have been left out.

The data provided are ideal for a fantastic classroom conversation about causation, causal mechanisms, and confounding variables.

\subsubsection*{Thinking outside the box}

In the second analysis, we work with the additional variable {\tt continent}.  The trends observed in the first analysis hold up in the second analysis (i.e., the difference declines over time in each of the continents).   However, there are additional considerations to be made, for example, the differences between the slopes across the continents (the analysis is available in the supplementary materials and not shown here).  We suggest additional explorations into the independence of the residuals and more advanced spatio-temporal patterns of literacy.

\subsection{Weather Data from NOAA Buoys}

The National Oceanic and Atmospheric Administration (NOAA) is the American federal agency in charge of collecting information and making decisions related to the oceans and the atmosphere.  Throughout North America, they supply weather stations which are located both along the coast as well as in the middle of the ocean (on buoys).  Among other variables, the weather stations collect information on wind, humidity, temperature, visibility, and atmospheric pressure.  The data is all publicly available on NOAA's website, \url{http://www.ndbc.noaa.gov/}.

\subsubsection*{Using dynamic data within a typical classroom}

Although the data do not constitute a random sample, they are very likely to be quite representative with respect to the difference in wind and air temperature at that location over the year.  In the supplementary files (not shown here), we used a paired analysis (i.e., subtract the two variables and treat them as a single variable) to find a confidence interval for the true difference in temperature between wind and air.  Also, we find a prediction interval for the difference in temperatures across individual measurements.

\subsubsection*{Thinking outside the box}

\begin{figure}[H]
\begin{center}
\begin{knitrout}
\definecolor{shadecolor}{rgb}{0.969, 0.969, 0.969}\color{fgcolor}
\includegraphics[width=\maxwidth]{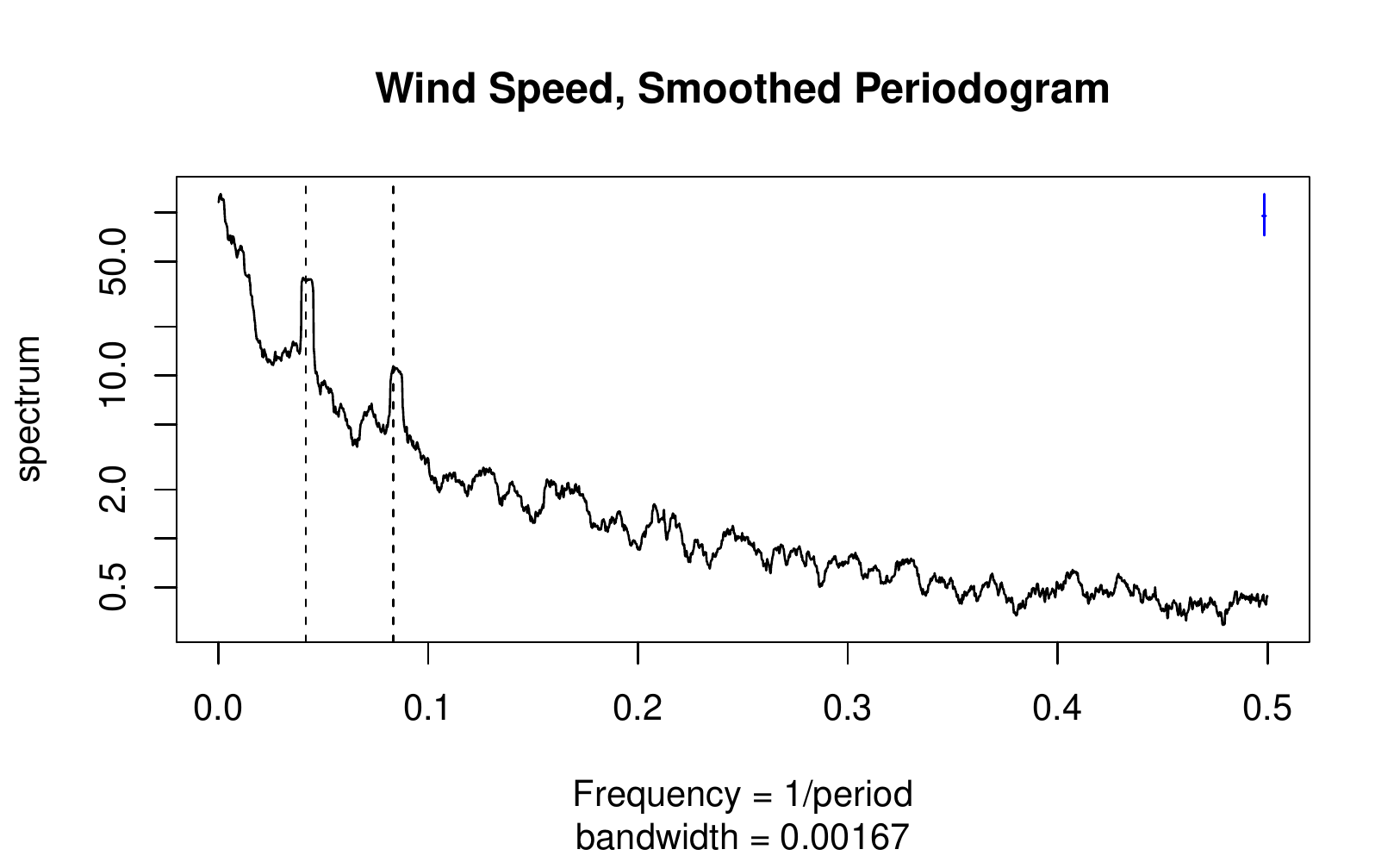} 

\end{knitrout}
\caption{\label{period} \footnotesize  A smoothed periodogram of Wind Speed for buoy $\#46025$ off the coast of Santa Monica.  The dashed lines are at 1/12 and 1/24, representing 12-hour and 24-hour periodicity.}
\end{center}
\end{figure}

Although a full analysis of the data would warrant multiple years of data (so as to understand yearly trends), we can estimate the spectral density of the time series using a smoothed periodogram (the data below represent measurements every hour for all of 2014).  In the smoothed periodogram (see Figure \ref{period}) the x-axis is the frequency (one over the period) and y-axis represents the correlation (normalized) between the cosine wave at that frequency and the time series.  We can see that wind speed has strong correlation at period 12 hours and period 24 hours.  A more sophisticated analysis or longer project could include collecting data from multiple buoys, extended years, and/or additional information on storms \url{https://www.ncdc.noaa.gov/stormevents/}.

\subsection{Other Dynamic Datasets}

There are myriad sources of dynamic data which are public and accessible.  We list a few additional resources here.

\subsubsection{Baseball Data}
Many students are interested in sports statistics.  In particular, statisticians and sabermatricians have worked over the years to compile datasets and to think carefully about statistical methods applied to baseball data. Today's datasets can answer questions like: What is the probability of a particular event happening given certain real time situations?

\begin{sloppypar}
\citet{albe:2010} provides curated seasonal batting data from 1871 to 2009 at \url{http://www.amstat.org/publications/jse/v18n3/mlb_batting.dat.txt}. With the passage of time, the examples of Derek Jeter and Alex Rodriguez as current players and batting trends up to 2009 are less relevant to students interested in sports, and what was once a current research question has become stagnant and ossified. However, \citet{albe:2010} provides sufficient detail about the collection of files on the Lahman Baseball Database (\url{http://www.seanlahman.com/baseball-archive/statistics/}  with data available as a .zip file which can be read directly into R \url{http://seanlahman.com/files/database/lahman-csv_2015-01-24.zip}) and tasks required to filter and merge the master and batting files that the research questions and teaching notes can be updated to reflect current players and trends.  Also see Jim Albert's blog {\em Exploring Baseball Data with R},  \url{https://baseballwithr.wordpress.com/}.
\end{sloppypar}

\subsubsection{Cherry Blossom Ten Mile Run Data}

Each year in April, the Cherry Blossom Ten Mile Run happens in Washington, DC (\url{http://www.cherryblossom.org/theraces/tenmile.php}).  The race results (including finishing time, name, age, hometown, and pace) are available from 1999 to the present (although admittedly, the results after 2012 are more difficult to scrape off the web due to a change in the format of how the results are now posted).  \citet{kapl:nola:2015} provide step-by-step instructions (\url{http://rdatasciencecases.org/}) for scraping the data and wrangling it into a format which can be used to investigate race results over time, by age, or by gender.

\subsubsection{Fatal Accidents and US Census Data}

The National Highway Traffic Safety Administration collects data on all fatalities suffered in motor vehicle traffic crashes.  The data is posted yearly and publicly as part of the Fatal Accident Reporting System, \url{http://www.nhtsa.gov/FARS}.  Each year is posted as a separate set of files, available in different formats.  For example, download the SAS data (readable into R via the {\em read.sas7bdat} function in the \verb;sas7bdat; package, the {\em sasxport.get} function in the \verb;Hmisc; package, or {\em read.SAS} in the \verb;haven; package) at \url{ftp://ftp.nhtsa.dot.gov/fars/2014/SAS/}.

Another dataset which is easily accessible is the census data (\url{http://factfinder2.census.gov/faces/nav/jsf/pages/index.xhtml}).  By looking at the {\em Decennial Census data} and the {\em Profile of General Population and Housing Characteristics}, students can study the ages and other demographic characteristics of the general population as compared to those individuals involved in fatal accidents\footnote{Original idea for this example from Laura Kapitula, Grand Valley State University.}.

\subsubsection{Climate Data}

\citet{witt:2013} describes using data in class to reveal important insights on climate for statistics students.  The data and analyses describe both decline of Arctic sea ice and global temperature increase.  The (static) data used in the analyses is available through {\em The Journal of Statistics Education}, \url{http://www.amstat.org/publications/jse/v21n1/witt.pdf}, but the authors also provide links and information about the original data sources which are dynamic.

Notably, realclimate.org provides a catalogue of many different types of climate data and relevant source information (\url{http://www.realclimate.org/index.php/data-sources/}).  \citet{witt:2013} also reports:

\begin{quote}
NASA's Global Change Master Directory is available at \url{http://gcmd.gsfc.nasa.gov/}. NOAA's National Climate
Data Center maintains an extensive data directory available at \url{http://www.ncdc.noaa.gov/oa/ncdc.html}. Yet another good climate data source is the Data Guide maintained by the National Center for Atmospheric Research Climate at \url{https://climatedataguide.ucar.edu/}.
\end{quote}

\subsubsection{Iowa Liquor Sales Data}

Iowa recently released an 800MB+ dataset containing all the weekly liquor sales from January 1, 2014 to the present (\url{https://data.iowa.gov/Economy/Iowa-Liquor-Sales/m3tr-qhgy}).  The data seems to be updated monthly.  Dan Nguyen has provided some scripting code and SQL analysis for downloading and wrangling the data (\url{https://gist.github.com/dannguyen/18ed71d3451d147af414}).  Because the data are available in a csv format, they can be read into R directly using read.csv from the data.iowa.gov URL.  The SQL code from the Nguyen's github site can be translated directly into R using the \verb;dplyr; package.  Note that the rectangular structure and comma delimited format of the data make it tempting to work with.  The challenge for this particular dynamic dataset comes with its size, which can make it unwieldy on many computers.

\subsubsection{Medicare Inpatient Charges Data}

\begin{sloppypar}
Medicare inpatient charge data is available at \url{https://www.cms.gov/Research-Statistics-Data-and-Systems/Statistics-Trends-and-Reports/Medicare-Provider-Charge-Data/Inpatient2013.html}
which links directly to a zipped csv file.  An analysis of the costs for Medicare Severity Diagnosis Related Group (MS-DRG) (i.e., the procedure) can be done over multiple different covariates.
\end{sloppypar}

\subsubsection{World Bank Data}

The R package {\tt wbstats} connects directly to the World Bank API, providing a trove of global economic data.  The package documentation and vignette provide an easy start for students to download basic data of their choosing, \url{https://cran.r-project.org/web/packages/wbstats/vignettes/Using_the_wbstats_package.html}.

\subsubsection{Energy Consumption Data}

The US Department of Energy collates energy usage at \url{http://www.eia.gov/}.  For example, at \url{http://www.eia.gov/coal/data.cfm} they provide complete data on ``coal production, consumption, exports, imports, stocks, mining, and prices."

\subsubsection{Government Survey Data}

Anthony Damico has compiled and documented a repository of data from many dozens of different government surveys \url{http://www.asdfree.com/}, including NHANES.  Additionally, Damico espouses the merits of reproducible analysis done in R with GitHub.  As one example, the General Social Survey (GSS) contains information on what Americans think of policies, issues, and priorities in the US (\url{http://gss.norc.org/}).

\subsection{Using the Examples}
We try to communicate to the students that there is information in most data sources.  We want to be wary and attentive to issues of experimental design and systematic biases.  However, we do not want to leave our students feeling stuck every time they encounter a dataset which has not been gathered from a large randomized trial.  Instead, we try to think of the pieces of the analysis that can elicit information which is interesting or possibly hypothesis generating.

Students at every level are ready to examine complicated relationships using tools which are accessible before taking multiple advanced statistics courses.  ``Unfortunately, many instructors teach the sections on data analysis as descriptive statistics, perhaps because this is what they experienced in their first course.  They emphasize the process of calculating numerical summaries and making graphical displays, rather than using these as tools to explore what the data are saying." \citep{notz:2015}

To keep our students from getting stuck, ``we must ask questions such as whether the data allow generalization to a larger population, whether their structure can be meaningfully described with the models we wish to fit, and whether important subgroups or individuals were excluded from the data.  Exceptions, anomalies, outliers, and subgroups are best recognized and understood in the context of the question being addressed." \citep{vell:2015}

\section{{ Conclusion}}

The few examples provided here give a glimpse into how to incorporate real and dynamic data into introductory  statistics and and courses beyond.  The Guidelines for Assessment and Instruction in Statistics Education (GAISE) College Report implores instructors to make one small change in our courses to keep up with changing technology and data \citep{gaise2015}.  In this manuscript we have provided tools for helping educators seamlessly incorporate dynamic data into a standard and modern introductory (or beyond) statistics course.  We also hope that statistics students will see the different ways of accessing data and feel empowered to gather more information on their own.  Our courses must remain relevant to both the student experience and the research community or we run the risk of become irrelevant.  Small steps will get us where we need to be.

\section{Acknowledgements}
\begin{sloppypar}
This work was supported by Project MOSAIC (NSF grant 0920350, Phase II: Building a Community around Modeling, Statistics, Computation, and Calculus).  Additionally, the data sources were found with help from many individuals.  For example, see Wes Stevenson's great blog on importing data into R (\url{http://statistical-research.com/importing-data-into-r-from-different-sources/}). An important step forward for accessing data is Jenny Bryan's R package \verb;googlesheets; (\url{http://blog.revolutionanalytics.com/2015/09/using-the-googlesheets-package-to-work-with-google-sheets.html}).  Many of the data sets were inspired by work by or conversations with Nick Horton, Ben Baumer, Gabe Chandler, Scott Grimshaw, Laura Kapitula, Danny Kaplan, Randy Pruim, Maddi Cowen, Ciaran Evans, Samantha Morrison, and Janie Neal.
\end{sloppypar}

\section{Appendix}

The Appendix contains technical details of interest to students who want to learn more about directly downloading data.  R Studio allows for the teacher to provide template R Markdown files which download data directly from the Internet into the R Studio environment.   Additionally, reproducible R Markdown files teach good science and analysis.  \citet{hort:baum:2015} discuss the merits of reproducible research as well as the ease of using R Markdown in the classroom.

As part of the supplementary materials (\url{https://github.com/hardin47/DynamicData}), we have provided R Markdown files giving both R code and related pedagogical commentary associated with different dynamic data examples for use in a statistics classroom.  The data have all been collected directly from outside sources - websites that are updated periodically (and with information the students can access on their own).

\subsection{What is an API?}

An Application Programming Interface (API) is a set of programming instructions (written in code, giving the appropriate algorithm) for downloading data from a website that contains - typically - vast amounts of data.  For example, Twitter has an API (\url{https://dev.twitter.com/overview/api}) which tells programmers how to access tweets (and related information) directly from the Twitter website.  At any level, but particularly useful at an introductory level, it is recommended to use an R (or other software) package or function that allows R to speak to the API in order to download the data.  For example, the R package \verb;wbstats; connects directly from R to the World Bank API and \verb;twitteR; provides an R interface to the Twitter API.  Though the examples provided do {\em not} generally rely on APIs or a related R interface, it is good to be aware of APIs (and to tell your students!) in order to greatly increase the sources of data available to you and your students.

\subsection{R Markdown}
The R Markdown files provided as supplementary materials make it easy for instructors and students to perform reproducible analyses on data collection through analysis and synthesis of the results.  The students get the important practice of tracking each and every step of their work.  Reproducible analysis has recently gotten a lot of press \citep{nytRR:2014,stod:2014}, and R Studio \citep{RStudio} has produced a user friendly format for combining R code with html (or \LaTeX) word processing that has been used with introductory statistics \citep{baumer:2014}.

R Markdown has a short learning curve and will be straightforward for your students to implement.  One of the key considerations is that when running an R Markdown file, the file does not pay attention to anything running locally.  R Markdown essentially restarts R, so the current state of your R session is totally irrelevant.  Recall that the purpose of R Markdown is to create reproducible files, so the Markdown file should run on any computer anywhere (regardless of the local environment). 

{\em Important note:}  Any package used in R Markdown needs to be installed prior to compiling the .Rmd file.  That is, for a line of code such as \verb;library(dplyr);, before running the Markdown file, \verb;install.packages("dplyr"); must be run from {\em within the console} one time (ever).

\subsection{Useful R functions}

Some important R {\em functions} (in {\em italics}) and \verb;packages; (in \verb;typeface;) that will help you and your students navigate importing and using data in R include:
\begin{itemize}
\item
\verb;dplyr; package for data wrangling in general; cheat sheet at \url{https://www.rstudio.com/wp-content/uploads/2015/02/data-wrangling-cheatsheet.pdf}
\item
The {\em glimpse} function in \verb;dplyr;
\item
\verb;tidyr; for converting between wide and long formats and for the very useful {\em extract\_numeric()} function  (or {\em readr::parse\_numeric()})
\item
\begin{sloppypar}
\verb;ggplot2; for faceted graphing (also, \verb;ggvis;) \citep{ggplot2}; cheat sheet at \url{https://www.rstudio.com/wp-content/uploads/2015/03/ggplot2-cheatsheet.pdf}
\end{sloppypar}
\item
\verb;openintro;  (or packages that come with the textbooks you use) which are great for pulling up any dataset from the text and building on it in class \citep{openintro}
\item
\begin{sloppypar}
\verb;mosaic; for consistent syntax and helpful functions used in introductory statistics \citep{mosaic}; cheat sheet at \url{https://cran.r-project.org/web/packages/mosaic/vignettes/MinimalR.pdf}
\end{sloppypar}
\item
\verb;googlesheets; for loading data directly from Google spreadsheets
\item
\verb;lubridate; if you ever need to work with any date fields
\item
\verb;stringr; for text parsing and manipulation
\item
\verb;rvest; for scraping data off the web; \verb;readxl; for reading excel data
\item
\verb;readr; (and \verb;fread; with {\em data.table}) for loading large datasets with default {\em stringsAsFactors = FALSE}
\item
\verb;tables; for nice looking summary tables.
\end{itemize}

\bibliographystyle{apa}
\bibliography{dyndata}

\begin{thebibliography}{28}
\newcommand{\enquote}[1]{``#1''}
\providecommand{\natexlab}[1]{#1}
\providecommand{\url}[1]{\texttt{#1}}
\providecommand{\urlprefix}{URL }
\expandafter\ifx\csname urlstyle\endcsname\relax
  \providecommand{\doi}[1]{doi:\discretionary{}{}{}#1}\else
  \providecommand{\doi}{doi:\discretionary{}{}{}\begingroup
  \urlstyle{rm}\Url}\fi
\providecommand{\eprint}[2][]{\url{#2}}

\bibitem[{Albert(2010)}]{albe:2010}
Albert J (2010).
\newblock \enquote{Baseball Data at Season, Play-by-Play, and Pitch-by-Pitch
  Levels.}
\newblock \emph{Journal of Statistics Education}, \textbf{18}(3).

\bibitem[{{American Statistical Association Undergraduate Guidelines
  Workgroup}(2014)}]{asa:2014}
{American Statistical Association Undergraduate Guidelines Workgroup} (2014).
\newblock \enquote{2014 curriculum guidelines for undergraduate programs in
  statistical science.}
\newblock \emph{Technical report}, American Statistical Association,
  Alexandria, VA.
\newblock
  \urlprefix\url{http://www.amstat.org/education/curriculumguidelines.cfm}.

\bibitem[{Baumer \emph{et~al.}(2014)Baumer, {\c{C}etinkaya-Rundel}, Bray, Loi,
  and Horton}]{baumer:2014}
Baumer B, {\c{C}etinkaya-Rundel} M, Bray A, Loi L, Horton N (2014).
\newblock \enquote{R Markdown: Integrating A Reproducible Analysis Tool into
  Introductory Statistics.}
\newblock \emph{Technology Innovations in Statistics Education}, \textbf{8}(1).
\newblock \urlprefix\url{http://escholarship.org/uc/item/90b2f5xh}.

\bibitem[{Brown and Kass(2009)}]{brow:kass:2009}
Brown EN, Kass RE (2009).
\newblock \enquote{What is Statistics?}
\newblock \emph{The American Statistician}, \textbf{63}(2), 105--110.

\bibitem[{Carver \emph{et~al.}(2016)Carver, Everson, Gabrosek, Rowell, Horton,
  Lock, Mocko, Rossman, Velleman, Witmer, and Wood}]{gaise2015}
Carver R, Everson M, Gabrosek J, Rowell GH, Horton N, Lock R, Mocko M, Rossman
  A, Velleman P, Witmer J, Wood B (2016).
\newblock \enquote{Guidelines for Assessment and Instruction in Statistics
  Education ({GAISE}) College Report.}
\newblock \emph{Technical report}, American Statistical Association.
\newblock
  \urlprefix\url{http://www.amstat.org/education/gaise/collegeupdate/GAISE2016\_DRAFT.pdf}.

\bibitem[{Cobb(1991)}]{cobb:1991}
Cobb GW (1991).
\newblock \enquote{Teaching Statistics: More Data, Less Lecturing.}
\newblock \emph{UME Trends}, pp. 3--7.

\bibitem[{Cobb(1992)}]{cobb:1992}
Cobb GW (1992).
\newblock \enquote{Teaching Statistics.}
\newblock \emph{In Lynn A. Steen (ed), Heeding the call for change: suggestions
  for curricular action (MAA Notes No. 22)}, pp. 3--43.

\bibitem[{Cobb(2007)}]{cobb:2007}
Cobb GW (2007).
\newblock \enquote{The Introductory Statistics Course: A {P}tolemaic
  Curriculum?}
\newblock \emph{Technology Innovations in Statistics Education}, \textbf{1}(1).
\newblock \urlprefix\url{https://escholarship.org/uc/item/6hb3k0nz}.

\bibitem[{Cobb(2011)}]{cobb:2011}
Cobb GW (2011).
\newblock \enquote{Teaching statistics: some important tensions.}
\newblock \emph{Chilean Journal of Statistics}, \textbf{2}(1), 31--62.

\bibitem[{{De Veaux} and Velleman(2015)}]{vell:2015}
{De Veaux} R, Velleman P (2015).
\newblock \enquote{Teaching Statistics Algorithmically or Stochastically Misses
  the Point: Why not Teach Holistically? (Online discussion of ``Mere
  Renovation is Too Little Too Late: We Need to Rethink Our Undergraduate
  Curriculum From the Ground Up," by {G}eorge {W} {C}obb, {\em {T}he {A}merican
  {S}tatistician}).}
\newblock \emph{The American Statistician}, \textbf{69}(4).

\bibitem[{Diez \emph{et~al.}(2012)Diez, Barr, and
  {\c{C}etinkaya-Rundel}}]{openintro}
Diez DM, Barr CD, {\c{C}etinkaya-Rundel} M (2012).
\newblock \emph{openintro: OpenIntro data sets and supplemental functions}.
\newblock R package version 1.4,
  \urlprefix\url{http://CRAN.R-project.org/package=openintro}.

\bibitem[{{GAISE College Group}(2005)}]{gaise}
{GAISE College Group} (2005).
\newblock \enquote{Guidelines for Assessment and Instruction in Statistics
  Education.}
\newblock \emph{Technical report}, American Statistical Association.
\newblock \urlprefix\url{http://www.amstat.org/education/gaise}.

\bibitem[{Gould(2010)}]{goul:2010}
Gould R (2010).
\newblock \enquote{Statistics and the Modern Student.}
\newblock \emph{International Statistical Review}, \textbf{78}(2), 297--315.

\bibitem[{Gould and {\c{C}etinkaya-{R}undel}(2013)}]{goul:cent:2013}
Gould R, {\c{C}etinkaya-{R}undel} M (2013).
\newblock \enquote{Teaching Statistical Thinking in the Data Deluge.}
\newblock In T~Wassong, D~Frischemeier, PR~Fischer, R~Hochmuth, P~Bender
  (eds.), \emph{Using Tools for Learning Statistics and Mathematics}, pp.
  377--391. Springer.

\bibitem[{Grimshaw(2015)}]{grim:2015}
Grimshaw S (2015).
\newblock \enquote{A Framework for Infusing Authentic Data Experiences Within
  Statistics Courses.}
\newblock \emph{The American Statistician}, \textbf{69}(4), 307--314.
\newblock \urlprefix\url{http://arxiv.org/abs/1507.08934}.

\bibitem[{Horton \emph{et~al.}(2015)Horton, Baumer, and
  Wickham}]{hort:baum:2015}
Horton NJ, Baumer B, Wickham H (2015).
\newblock \enquote{Setting the stage for data science: integration of data
  management skills in introductory and second courses in statistics.}
\newblock \emph{CHANCE}, \textbf{28}(2), 40--50.
\newblock \urlprefix\url{http://arxiv.org/abs/1502.00318}.

\bibitem[{Johnson(2014)}]{nytRR:2014}
Johnson G (2014).
\newblock \enquote{New Truths That Only One Can See.}
\newblock
  \urlprefix\url{http://www.nytimes.com/2014/01/21/science/new-truths-that-only-one-can-see.html}.

\bibitem[{Kaplan and Nolan(2015)}]{kapl:nola:2015}
Kaplan D, Nolan D (2015).
\newblock \enquote{Modeling Runners' Times in the Cherry Blossom Race.}
\newblock In \emph{Data Science in {R}: A Case Studies Approach to
  Computational Reasoning and Problem Solving}, pp. 45--104. CRC Press.

\bibitem[{Kuiper and Sturdivant(2015)}]{stur:kuip:2015}
Kuiper S, Sturdivant RX (2015).
\newblock \enquote{Using Online Game-Based Simulations to Strengthen Students'
  Understanding of Practical Statistical Issues in Real-World Data Analysis.}
\newblock \emph{The American Statistician}, \textbf{69}(4), 354--361.

\bibitem[{Notz(2015)}]{notz:2015}
Notz W (2015).
\newblock \enquote{Vision or Bad Dream? (Online discussion of ``Mere Renovation
  is Too Little Too Late: We Need to Rethink Our Undergraduate Curriculum From
  the Ground Up,'' by {G}eorge {W} {C}obb, {\em {T}he {A}merican
  {S}tatistician}).}
\newblock \emph{The American Statistician}, \textbf{69}(4).

\bibitem[{Pruim \emph{et~al.}(2014)Pruim, Kaplan, and Horton}]{mosaic}
Pruim R, Kaplan D, Horton N (2014).
\newblock \emph{mosaic: {Project MOSAIC} (mosaic-web.org) Statistics and
  Mathematics Teaching Utilities}.
\newblock R package version 0.9-1-3,
  \urlprefix\url{http://CRAN.R-project.org/package=mosaic}.

\bibitem[{{RStudio Team}(2015)}]{RStudio}
{RStudio Team} (2015).
\newblock \emph{RStudio: Integrated Development Environment for R}.
\newblock {RStudio, Inc.}, Boston, MA.
\newblock \urlprefix\url{http://www.rstudio.com/}.

\bibitem[{Stodden \emph{et~al.}(2014)Stodden, Leisch, and Peng}]{stod:2014}
Stodden V, Leisch F, Peng RD (eds.) (2014).
\newblock \emph{Implementing Reproducible Research}.
\newblock Chapman and Hall / CRC Press.

\bibitem[{Wickham(2014)}]{wick:2014}
Wickham H (2014).
\newblock \enquote{Tidy Data.}
\newblock \emph{Journal of Statistical Software}, \textbf{59}(10).
\newblock \urlprefix\url{http://www.jstatsoft.org/v59/i10/}.

\bibitem[{Wickham and Sievert(2016)}]{ggplot2}
Wickham H, Sievert C (2016).
\newblock \emph{{ggplot2:} Elegant Graphics for Data Analysis}.
\newblock Springer New York.
\newblock \urlprefix\url{http://had.co.nz/ggplot2/book}.

\bibitem[{Witt(2013)}]{witt:2013}
Witt G (2013).
\newblock \enquote{Using Data from Climate Science to Teach Introductory
  Statistics.}
\newblock \emph{Journal of Statistics Education}, \textbf{21}(1).

\bibitem[{{Workgroup on Undergraduate Statistics}(2000)}]{asa-undergrad}
{Workgroup on Undergraduate Statistics} (2000).
\newblock \enquote{Guidelines for Undergraduate Statistics Programs,
  \url{http://www.amstat.org/education/curriculumguidelines.cfm}, accessed
  {A}ugust 18, 2013.}
\newblock \emph{Technical report}, American Statistical Association.

\bibitem[{Zhu \emph{et~al.}(2013)Zhu, Hernandez, Mueller, Dong, and
  Forman}]{zhu:2013}
Zhu Y, Hernandez LM, Mueller P, Dong Y, Forman MR (2013).
\newblock \enquote{Data Acquisition and Preprocessing in Studies on Humans:
  What is Not Taught in Statistics Classes?}
\newblock \emph{The American Statistician}, \textbf{67}, 235--241.

\end{thebibliography}

\end{document}